\documentclass[%
 reprint,
superscriptaddress,
 amsmath,amssymb,
 aps,
pra,
showkeys
]{revtex4-2}

\usepackage{float}
\usepackage{graphicx}
\usepackage{dcolumn}
\usepackage{bm}
\usepackage{upgreek}
\usepackage{xcolor} 
\usepackage{soul,xcolor}
\usepackage{amsmath}
\usepackage{fancyhdr}
\setstcolor{orange}

\begin{document}

\preprint{APS/123-QED}

\title{A comparison of continuous and pulsed sideband cooling on an electric quadrupole transition}

\author{Evan C. Reed}
\email{er118@duke.edu}
\affiliation{Duke Quantum Center, Duke University, Durham, NC 27701, USA}
\affiliation{Department of Electrical and Computer Engineering, Duke University, Durham, NC 27708, USA}
\author{Lu Qi}
\affiliation{Duke Quantum Center, Duke University, Durham, NC 27701, USA}
\affiliation{Department of Electrical and Computer Engineering, Duke University, Durham, NC 27708, USA}
\author{Kenneth R. Brown}
 \email{kenneth.r.brown@duke.edu}
 \affiliation{Duke Quantum Center, Duke University, Durham, NC 27701, USA}
\affiliation{Department of Electrical and Computer Engineering, Duke University, Durham, NC 27708, USA}
\affiliation{Departments of Physics and Chemistry, Duke University, Durham, NC 27708, USA}

\date{\today}

\begin{abstract}
Sideband cooling enables preparation of trapped ion motion near the ground state and is essential for many scientific and technological applications of trapped ion devices.
Here, we study the efficiency of continuous and pulsed sideband cooling using both first- and second-order sidebands applied to an ion where the motion starts outside the Lamb-Dicke regime.
We find that after optimizing these distinct cooling methods, pulsed and continuous cooling achieve similar results based on simulations and experiments with a $^{40}$Ca$^+$ ion.
We consider optimization of both average phonon number $\overline{n}$ and population in the ground state.  
We also demonstrate the disparity between $\overline{n}$ as measured by the sideband ratio method of trapped ion thermometry and the $\overline{n}$ found by averaging over the ion's motional state distribution.


\end{abstract}

\maketitle

\section{\label{intro} Introduction}
Resolved sideband cooling (SBC) allows for preparation of a trapped ion register near the ground state of motion through the application of laser light resonant with a red sideband of a spectrally narrow electronic transition \cite{Diedrich1989}.
Due to reduced motional decoherence and the elimination of Doppler shifts, the cooling technique has laid the foundation for the ongoing development of trapped ion based technologies such as quantum computers \cite{Monroe1995, Bruzewicz2019}, simulators \cite{Monroe2021}, sensors \cite{Wolf2021}, quantum repeaters \cite{Krutyanskiy2023}, and atomic clocks \cite{Brewer2019}.
Beyond the technological applications, resolved sideband cooling enables the spectroscopy and control of molecular ions \cite{rugango2015sympathetic, Qi2023, alighanbari2018rotational, Chou2017, Wolf2016}, experimental demonstrations of quantum entanglement \cite{Blatt_Wineland_2008, Lin2020}, cooling of a levitated nano-particle \cite{PhysRevLett.123.153601}, and molecular and atomic ion enabled searches for new physics beyond the Standard Model \cite{Budker2022, Kajita2009}. 

SBC is of particular importance to the operation of trapped ion quantum computers. 
These devices rely on two-qubit gates, which achieve optimal performance when the ions are cooled to near the ground state of motion.
Ground state cooling takes up a majority of the amount of the time required to prepare the qubit register and a significant amount of the time needed to run a quantum circuit. 
In certain trapped ion processor architectures, such as the quantum charge coupled device (QCCD) architecture \cite{Kielpinski2002CCD}, ion shuttling necessitates frequent recooling, which can take anywhere between 25-68\% of the processing time of a given quantum circuit \cite{kaushal2020shuttling, Moses2023}. 
Moreover, as quantum computers continue to scale and circuit depths increase, mid-circuit cooling will become more vital in order to reinitialize the motional state after measurement or to limit the effects of motional decoherence due to ion heating \cite{Turchette2000heating, Deslauriers2006, PRXQuantum.3.010334}. 
Therefore, the need to cool to the motional ground state is a significant source of latency in a quantum algorithm on a trapped ion processor, and efficient SBC is needed to minimize this latency.
However, despite this need for efficient ground state cooling, the two means of implementing SBC, pulsed and continuous, have not been characterized with respect to each other.

Electromagnetically induced transparency (EIT) cooling is another technique that can be used for cooling to close to the ground-state of motion \cite{Morigi2000, Roos2000, Lechner2016}.
However, EIT cooling often requires an additional laser frequency in order to generate the coherent effect, which introduces additional experimental complexity.
Furthermore, EIT cooling has a relatively high cooling limit, so it is often followed by a short period of SBC to completely prepare the ground-state of motion \cite{ChaoThesis}.
For these reasons, many trapped ion experiments solely rely on SBC to prepare the motional ground state, and therefore, we consider here only SBC starting from a post-Doppler cooling motional distribution. 
Moreover, while pulsed Raman SBC is frequently utilized with hyperfine qubits \cite{Rasmusson2021, Wu2023} and continuous Raman SBC was recently demonstrated \cite{Che2017}, here we limit our discussion to SBC on an electric quadrupole (E2) transition.

In this report, we investigate the comparative cooling efficiencies of continuous and pulsed SBC on an E2 transition as characterized by the rate of ground state population $P_{n=0}(t)$ and the reduction of average phonon number $\overline{n}$.
We numerically simulate both techniques in order to find optimized parameters to ensure a fair comparison.
We also do the same for multi-order SBC to investigate the performance of pulsed and continuous SBC when cooling is performed on higher order sidebands. 
To confirm the validity of the numerical simulations, we verify the predictions experimentally by implementing optimized cooling on a trapped $^{40}$Ca$^{+}$ ion.
Our study predominantly centers on cooling starting from outside the Lamb-Dicke (LD) regime as this reflects the status of our experimental apparatus, which has a low axial frequency to be compatible with a dipole-phonon quantum logic experiment \cite{Campbell_Hudson_2020, mills2020dipole, Qi2023}.
We find that, when optimized, pulsed and continuous SBC produce similar results with pulsed cooling having a slight advantage particularly when implemented with higher order sidebands.
Additionally, we find that for continuous cooling one is able to optimize for either fast population of the ground state or fast cooling of $\overline{n}$. 
The appropriate optimization would depend on the application. 

This paper is organized in the following way.
Section \ref{sec:background} provides a brief overview of resolved SBC and touches on the differences between the continuous and pulsed techniques. 
In Section \ref{sec:numericalSim}, we examine the numerical simulation methods.  
We also discuss how we convert our simulation results to a ubiquitous method of ion thermometry, the sideband ratio (SBR) method, in order for the simulations to be comparable to experiments.
In Section \ref{sec:exp}, we describe the details of our experimental set-up. 
Finally, in Section \ref{sec:results}, we present the results of both the numerical simulations and the experiments, and Section \ref{sec:discuss} considers the pros and cons of the two approaches to SBC. 
Cooling performance from within the LD regime is shown in the Appendix along with descriptions of the optimization of multi-order continuous cooling parameters and the experimental calibration of the light shift and repumping laser parameters for continuous cooling.

\begin{figure}[t]
    \centering
    \includegraphics[width=\linewidth]{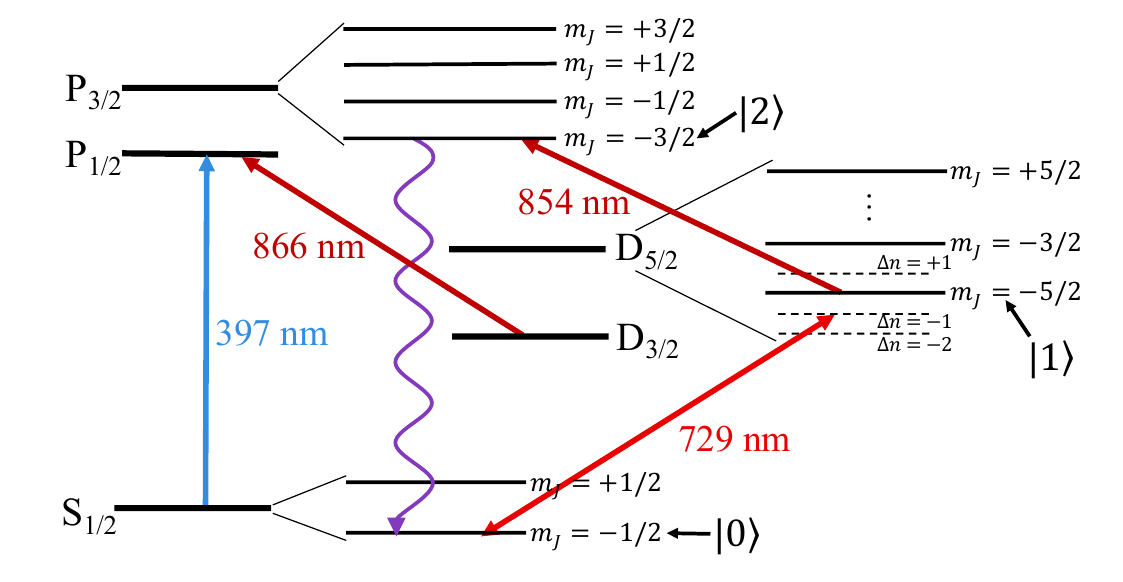}
    \caption{An energy level diagram of $^{40}$Ca$^+$ (not to scale). 
    Doppler cooling is performed along the 397\,nm transition, and the 866\,nm laser repumps leakage from the P$_{1/2}$ state to the D$_{3/2}$ and closes the Doppler cooling cycle. 
    Sideband cooling is performed on the red motional sidebands ($\Delta n < 0$) of the 729\,nm transition ($|0\rangle \leftrightarrow |1\rangle$), and quenching/repumping of the excited state is performed by the 854\,nm laser ($|1\rangle \rightarrow |2\rangle$). 
    The auxiliary excited state $|2\rangle$ is short lived and quickly decays to repopulate the ground electronic state $|0\rangle$.}
    \label{fig:energylevels}
\end{figure}
\begin{figure}[t]
    \centering
    \includegraphics[width=\linewidth]{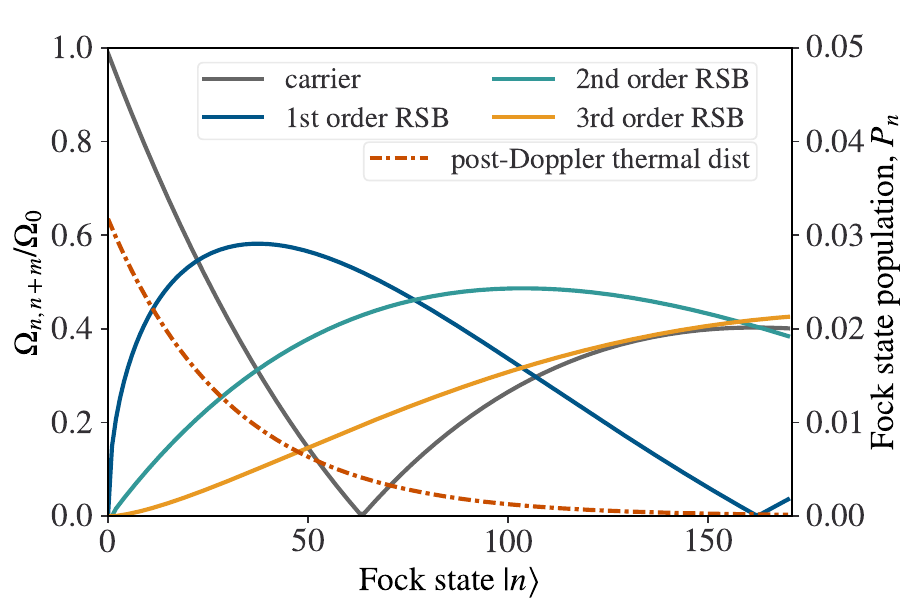}
    \caption{The relative strengths of the carrier and RSBs of the 729\,nm transition along with the post-Doppler motional state population distribution (thermal distribution) determined by the experimental parameters. 
    The solid lines show the normalized Rabi frequencies for the carrier and red sidebands (left y-axis) while the dash-dotted line give the motional state distribution (right y-axis). }
    \label{fig:RabiStrengths}
\end{figure}

\section{\label{sec:background} Background}
After an ion is initially trapped, Doppler cooling reduces the ion temperature to the to the milli-Kelvin temperature regime in few milliseconds.
In this process, a laser is red detuned from resonance with a fast decaying (electric dipole, or E1) transition by about half the magnitude of the transition linewidth \cite{Wineland1978}. 
In Ca$^{+}$, as shown in Fig.~\ref{fig:energylevels}, this is performed on the S$_{1/2} \rightarrow$ P$_{1/2}$ transition with an additional laser on the D$_{3/2} \rightarrow$ P$_{1/2}$ transition to repump branching outside of the Doppler cooling cycle.
However, Doppler cooling in a Paul trap has a fundamental lower limit to the attainable temperature: $\overline{n}=\Gamma/(2\omega_t)$, where $\Gamma$ is the transition linewidth and $\omega_t$ is the trap secular frequency.
For experimental systems with shallow trapping potentials, this lower limit may not put the ion in the Lamb-Dicke  (LD) regime, which is characterized by 
\begin{equation}
    \eta^2(2n+1)\ll 1 \; ,
    \label{eq:LDregime}
\end{equation}
where
\begin{equation}
    \eta=(2\pi/\lambda)\cos{(\theta)}\sqrt{\hbar/2m\omega_t}
    \label{eq:LDparam}
\end{equation}
is the LD parameter.
Here, $\theta$ is the angle between the laser propagation and the mode of secular oscillation characterized by frequency $\omega_t$.
In this regime, the spatial distribution of the ion's wave-packet becomes smaller than the wavelength of the applied laser field and higher order transitions between motional states ($\Delta n > 1$) become strongly suppressed.
Many trapped ion experiments and quantum operations require either ground state preparation or preparation in the LD regime near the motional ground state \cite{Cirac_Zoller_1995, Sorensen_Molmer_1998}.
Therefore, a second stage of cooling beyond Doppler cooling is often necessary.

Taking the states of a Ca$^{+}$ ion as shown in Fig.~\ref{fig:energylevels} as an example, the process of resolved sideband cooling is as follows: 
(1) prepare the electronic ground state $|0\rangle=|S_{1/2},m_J=-1/2\rangle$ via optical pumping. 
(2) Excite the red sideband of a narrow linewidth transition $(\Gamma_{01} \ll \omega_t)$ to a metastable excited state $|1\rangle=|D_{5/2},m_J=-5/2\rangle$  with a frequency-locked, narrow linewidth cooling laser: $\omega_{729} =\omega_{01}+m\omega_{t}$, where $m=\Delta n<0$.
(3) Quench the population of the metastable excited state by exciting to a short lived auxiliary excited state $|2\rangle=|P_{3/2},m_J=-3/2\rangle$, which will quickly decay back to the electronic ground state $|0\rangle$. 
(4) Repeat steps 2 and 3 until cooled to the ground state while periodically repeating step 1 as needed to maintain a closed cooling cycle. 

In the process of pulsed SBC, the red sideband (RSB) of $|0\rangle \leftrightarrow|1\rangle$ is excited with a precisely timed pulse to maximally transfer population to the metastable excited state.
Next, a quenching pulse is applied for a duration on the order of tens of microseconds. 
If the cooling pulse is implemented on the first order RSB ($m=-1$), then one quantum of motion is removed from the distribution of motional states; a pulse tuned to the second order RSB removes two phonons, and so on. 
Moreover, transitions between Fock states ($|n\rangle\rightarrow|n-1\rangle$) have coupling strengths that depend on the motional quantum number, as seen in Eq.~\ref{eq:RabiFreq}, and thus cooling pulses can be timed to target specific regions of the motional state distribution \cite{Qi2023}.

In continuous SBC, however, the quenching laser remains on throughout the cooling cycle and provides a constant coupling between the metastable and the auxiliary excited states \cite{Diedrich1989, Marzoli1994, Morigi1997, Sawamura2008, Joshi2019, Kulosa2023}. 
It is convenient to consider this coupling between these states, $|1\rangle$ and $|2\rangle$, in terms of the energy eigenstates of the coupling Hamiltonian (i.e., the dressed basis): 
$\{|1\rangle, |2\rangle\} \rightarrow \{|\psi_+\rangle, |\psi_-\rangle\}$. 
Then, in the limit that the coupling is characterized by a small saturation parameter, the three level system ($|0\rangle,|1\rangle,|2\rangle$) can be reduced to an effective two level system ($|0\rangle,|\psi_{+/-}\rangle$) with an effective linewidth given by \cite{Marzoli1994}
\begin{equation}
    \Gamma_{\mathrm{eff}}=\Gamma_{20} \frac{\left(\Omega_{12} / 2\right)^2}{\left[\left(\Gamma_{20}+\Gamma_{12}\right) / 2\right]^2+\delta_{12}^2} \; ,
    \label{eq:effectiveLinewidth}
\end{equation}
where $\Gamma_{if}$, $\Omega_{if}$, and $\delta_{if}$ are the natural linewidth, Rabi frequency, and detuning, respectively, between an initial and final state.
Note that $\Gamma_{\mathrm{eff} }\ll \omega_z$ must be maintained for the resolved sideband regime to hold. 
This new effective linewidth, which in general satisfies $\Gamma_{\mathrm{eff}} \gg \Gamma_{01}$, allows for fast decay back to the ground electronic state after excitation. 
Additionally, this coupling shifts the excited state from the previous $|0\rangle \leftrightarrow|1\rangle$ resonance by an amount given by 
\begin{equation}
    \Delta=\delta_{01}-\delta_{12} \frac{\left(\Omega_{12} / 2\right)^2}{\left[\left(\Gamma_{20}+\Gamma_{12}\right) / 2\right]^2+\delta_{12}^2} \; .
     \label{eq:LightShift}
\end{equation}
This is the AC Stark shift, or light shift. 
Also note that the effective linewidth and light shift depend primarily on the Rabi frequency $\Omega_{12}$ and detuning $\delta_{12}$ of the repumping laser \cite{Marzoli1994}. 

Therefore, while pulsed and continuous SBC are similar in concept $-$ excite the red sideband to remove a phonon $-$ they are quite different in practice. 
Efficient pulsed cooling relies on a set of time-tailored cooling pulses to maximally and coherently shift the population from the electronic ground state to the metastable state removing a phonon in the process. 
Meanwhile, efficient continuous cooling is an entirely incoherent process that relies on a repumping laser with carefully tuned Rabi frequency and detuning in order to reduce a long lived metastable state to an effective fast decaying excited state while maintaining resolved sidebands.

\section{\label{sec:numericalSim} Numerical Simulations}
As a means of both comparing the relative cooling efficiencies of pulsed and continuous SBC as well as finding optimum values for the following experiments, we numerically simulate the two cooling techniques. 
For continuous cooling, we directly evolve the interaction Hamiltonian with the quantum master equation, which models the interaction of a trapped, three-level ion with two light fields and bosonic motional modes without assuming the LD approximation. 
For pulsed cooling, we utilize a recently proposed and experimentally demonstrated graphic theoretic method of SBC pulse optimization \cite{Rasmusson2021}. 
In these simulations, we assume that the post-Doppler distribution of motional states is a thermal distribution  \cite{Javanainen_Stenholm_1981}.
Due to the shallow trapping potential, the $\overline{n}$ of thermal distribution in the axial mode is $30.5 \pm 2.2$ phonons as determined by a fit of the Rabi flopping on the carrier transition.
The post-Doppler thermal distribution of motional states is shown in Fig.~\ref{fig:RabiStrengths}.
To measure the motional excitation of the trapped ion after SBC, we use the sideband ratio (SBR) method of thermometry. 
In the subsection below, we highlight the disparity between average phonon number as measured by the SBR method, $\overline{n}_{SBR}$, and the average phonon number that one would expect from the results of numerical simulation, $\overline{n}_{NS}$.
We discuss how we convert our simulation results to the SBR method to allow the simulations to be comparable to the experimental results.

\subsection{Continuous sideband cooling}
Consider a three-level atom with states $|0\rangle$, $|1\rangle$, and $|2\rangle$ as described in the previous section. 
One could reasonably use the quantum Rabi model to describe the interaction of light fields and atomic states $|i\rangle \leftrightarrow |f\rangle$ as parameterized by the ground state Rabi frequency $\Omega^{if}_0$, detuning $\delta_{if}$, and LD parameter $\eta_{if}$:
\begin{multline}
    H_{i\leftrightarrow f} = \frac{\hbar\Omega^{if}_0}{2}\bigg[1-\eta_{if}^2 \bigg(a^{\dag}a + \frac{1}{2}\bigg)\sigma_-  e^{i \delta_{if} t} \\
    - i\eta_{if} a^{\dag}\sigma_- e^{i(\delta_{if} - \omega_t)t} 
    + i\eta_{if} a^{\dag}\sigma_+ e^{-i(\delta_{if} + \omega_t)t}
    \bigg] + H.c. \; ,
    \label{eq:JChamiltonian}
\end{multline}
where $\sigma_- = |i\rangle\langle f|$ is the lowering operator for the atom and $a$ is the annihilation operator for the motional mode.
The problem with the quantum Rabi model in this case, however, is the lack of generality caused by the  assumption that the ion is in the LD regime (Eq.~\ref{eq:LDregime}). 
At the post-Doppler average phonon number of $30.5$ phonons with a secular frequency of $415$\,kHz, approximately $50\%$ of the motional state distribution of our trapped Ca$^{+}$ ion meets the criteria $\eta^2(2n+1)\geq 1 $ and thus falls outside of the LD regime. 
To reclaim generality, we can describe how a Fock state $|n\rangle$ couples to a Fock state $|n+m \rangle$ by the following expression for Rabi frequency \cite{Wineland1998}:
\begin{multline}
    \Omega^{if}_{n,n+m}=\Omega^{if}_0 \left\langle n+m\left|e^{i \eta_{if}\left(a+a^{\dagger}\right)}\right| n\right\rangle \\
    =\Omega^{if}_0 \exp \left(-\frac{\eta_{if}^2}{2}\right) \left(\frac{n_{<} !}{n_{>} !}\right)^{1 / 2} \eta^{|m|} L_{n_{<}}^{|m|}\left(\eta_{if}^2\right) \; ,
    \label{eq:RabiFreq}
\end{multline}
where the Laguerre polynomials are given by 
\begin{equation}
    L^{m}_n(x) = \sum^n_{k=0}(-1)^k \binom{n+m}{n-k} \frac{x^k}{k!} \; .
    \label{eq:Laguerre}
\end{equation}

Therefore, we used this Rabi frequency matrix element to construct a coupling matrix for each carrier and sideband transition that describes how every Fock state couples to every other Fock state. 
We can thus rewrite the Hamiltonian as 
\begin{multline}
    H_{i \leftrightarrow f} = \frac{\hbar\Omega^{if}_0}{2} \sigma_{-}  \bigg[  C_0 e^{i \delta_{if} t}- i  C_{-1} e^{i\left(\delta_{if}+\omega_t\right) t}\\ 
    -i  C_{+1} e^{i\left(\delta_{if}-\omega_t\right) t} + ... \bigg] + H.c. \: ,
    \label{eq:compHamiltonian}
\end{multline}
where $C_{\pm m}$ is the coupling matrix for the carrier ($m=0$) or sidebands ($m\neq0$).
This method of constructing the Hamiltonian allows one to add arbitrarily higher sidebands orders with ease and does not implement the LD approximation. 
Using this Hamiltonian, one can describe the evolution of the system using the quantum master equation: 
\begin{equation}
    \dot{\rho}=-\frac{i}{\hbar}[H, \rho]+\sum_k \Gamma_k\left\{L_k \rho L_k^{\dagger}-\frac{1}{2}\left(L_k L_k^{\dagger} \rho + \rho L_k L_k^{\dagger}\right)\right\} \,  ,
    \label{eq:masterEq}
\end{equation}
where $\rho$ is the density matrix describing the internal and motional states of the ion and $L_k$ are the Lindbladian operators, which describe non-unitary interactions with the environment such as excited state decay and motional heating. 
The Lindbladian operators used in these simulations are provided in the Appendix.
\begin{figure}[t]
    \centering
    \includegraphics[width=\linewidth]{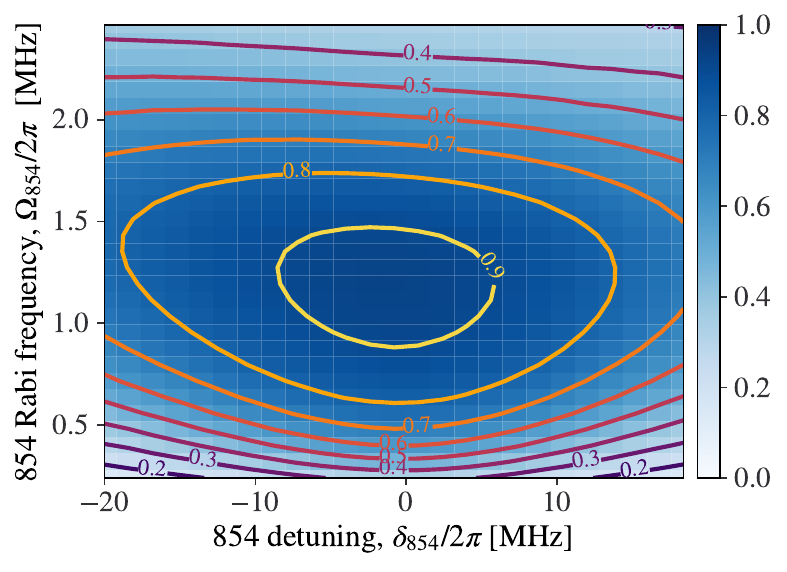}
    \caption{A 2D heat map of the ground motional state population as a function of repumping laser Rabi frequency and detuning for single order continuous 
    SBC at a total cooling time of 2\,ms. The contour lines are included to help guide the eye as to the boundaries of the population decades. 
    This plot shows that the best cooling happens when the repumping laser is close to resonance and with small Rabi frequency.}
    \label{fig:2Dsim}
\end{figure}

\begin{figure}[t]
    \centering
    \includegraphics[width=\linewidth]{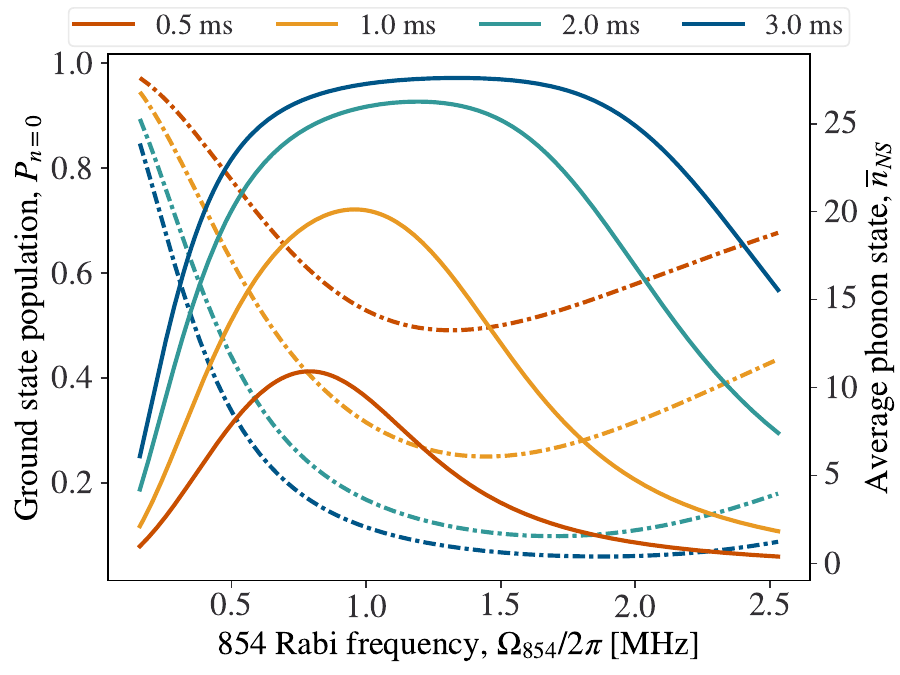}
    \caption{A slice of the 2D plot in Fig.~\ref{fig:2Dsim} at $\delta_{854}=0$\,MHz depicting various cooling times. 
    Here, the solid lines show the ground motional state population (left y-axis) while the dash-dotted lines show the average phonon number (right y-axis). 
    This plot demonstrates how the optimum value of $\Omega_{854}$ ($\equiv \Omega^{12}_{0}$) depends on the choice to either most efficiently populate the ground motional state or most efficiently lower the average phonon number. 
    It also shows that the optimum $\Omega_{854}$ changes as a function of cooling time.}
    \label{fig:minNbar_maxPn=0}
\end{figure}

The numerical simulations of continuous SBC were performed using the QuTiP master equation solver with parameters that reflect our experimental conditions \cite{qutip}. 
For a fixed cooling laser Rabi frequency, $\Omega^{01}_0 (\equiv \Omega_{729})$, we found the optimum repumping laser Rabi frequency, $\Omega^{12}_0 (\equiv \Omega_{854}$), and detuning, $\delta_{12} (\equiv \delta_{854})$, that allow for the most efficient cooling. 
The results of the continuous SBC optimization for cooling on the first order sideband are shown in Figs. \ref{fig:2Dsim} and \ref{fig:minNbar_maxPn=0}.
In Fig.~\ref{fig:2Dsim}, we can see that cooling is most efficient when the repumping laser is close to resonance with the $|1\rangle \leftrightarrow |2\rangle$ transition.
While the efficiency of the cooling is loosely dependent on detuning, the Rabi frequency requires more precision to attain efficient cooling.
However, from Fig.~\ref{fig:minNbar_maxPn=0} we observe that the optimal value of $\Omega^{12}_0$ changes as a function of cooling time, and from this we infer a dependence of the optimal Rabi frequency on the distribution of motional states.
Therefore, one could possibly further optimize continuous SBC by introducing time-dependent repumping Rabi frequency, although this was not implemented in this study. 
It is also important to note that the repumping  Rabi frequency that maximizes the ground state population is not the same as the Rabi frequency that minimizes the average phonon number. 
In the following single-order SBC experiments, we use the repumping Rabi frequency that maximizes the ground state population over minimizing $\overline{n} $ as determined by simulation.
At a cooling time of $3$\,ms, we found that the optimal repumping laser Rabi frequency is $1.34$\,MHz, which in our system is on the order of $1$\,uW of optical power. 
This value is consistent with a recently published analysis of continuous SBC \cite{Kulosa2023}. 
We also find good agreement between our simulations and the results of Ref. \cite{Joshi2019}.

The optimization of the repumping laser parameters becomes a more difficult task when one implements cooling on multiple sideband orders. 
For simplicity, we consider cooling on just the second and first order sidebands, consecutively.
In this case, in addition to Rabi frequency and detuning, one must also optimize the ratio of the total cooling time spent on the second order sideband vs the first order, which was done here with a brute force approach.
Moreover, while the first order sideband was optimized for maximizing the ground state population, cooling on the second order was optimized for minimizing the average phonon number. 
This is because the primary motivation for cooling on higher order sidebands, in addition to the removal of more than one phonon per cooling cycle, is to target the motional state population where the first order sideband is weak ($n>80$) and to prevent population trapping above where $\Omega_{n,n-1}$ goes to zero. 
These regimes can be seen in Fig.~\ref{fig:RabiStrengths}.
A more in-depth discussion on the optimization of multi-order continuous SBC can be found in the Appendix.

\subsection{Pulsed sideband cooling} 
The challenge in implementing effective pulsed SBC is to design pulse trains that will efficiently remove phonons from the initial post-Doppler thermal distribution. 
Traditionally, pulse trains are designed to be a series of pulses that target states in different regions of the motional state distribution, i.e., short pulses for high energy motional states and long pulses for low energy states.  
For a known initial motional state, $|n\rangle$, this would be a sequence of $n$ dependent $\pi$ pulses. 
We previously used a traditional pulse train design in our experimental set-up as described in Ref. \cite{Qi2023}. 
However, manually designed pulse trains may not be the most efficient means of cooling a trapped ion system. 
Here, we use a graph theoretic method of modeling pulsed SBC to determine the most efficient set of pulses for our system following Ref. \cite{Rasmusson2021}. 
This method models the Fock states as nodes on a graph, which is described mathematically as a vector of Fock state populations: $\overrightarrow{p} = \{p(n=0), p(1),...,p(n_{\textrm{max}})$.
Connections between nodes model the probability of cooling, $b_n (\tau) = \sin^2(\Omega_{n,n-1}\tau/2)$, and feedback from a node back into itself represents the probability of not cooling, $a(\tau)=1-b(\tau) = \cos^2(\Omega_{n,n-1}\tau/2)$. 
From these probabilities, a pulse matrix, $W(\tau)$, can be constructed, which acts on the Fock state population vector to describe how a cooling pulse affects the motional state distribution: $W^N(\tau)\overrightarrow{p_i} = \overrightarrow{p_f}$, where $N$ is the total number of pulses.
With this model, one can use a computational minimizing function to find the shortest pulse time $\tau$ that gives the lowest average phonon number $\overline{n}$.
In this computational minimization, we assume the value of $\tau$ to be the same for all pulses. 
It is possible to optimize $\tau$ for each pulse, but it was observed in Ref. \cite{Rasmusson2021} that this provides no advantage with respect to cooling efficiency.
We observe this in our simulations as well. 
Here, the pulse time $\tau$ assumes a constant cooling laser Rabi frequency. 
Therefore, only square pulses are used in this implementation of pulsed SBC. 
Additionally, in multi-order SBC the optimum value of $\tau$ may differ between sideband orders, although, as previously mentioned, we do not consider sidebands higher than the second order.

Recall that in our continuous SBC analysis a difference in optimal parameters was found when optimizing for $\overline{n}$ minimization versus $P_{n=0}$ maximization.
Therefore, we tried altering the computational optimizer to maximize $P_{n=0}$ rather than minimize $\overline{n}$ but found little difference in the pulsed cooling results contrary to continuous SBC. 
Also note that the graphic theoretic method currently does not account for practical deviations from an ideal cooling pulse such as off-resonant coupling, ion heating, or other sources of decoherence.
In spite of this, however, when we evolved a sample pulse train with the master equation solver ($50$ pulses starting from $\overline{n}_i=7$), we found that the QuTiP solver and the graph theoretic approach produced consistent results. 
Note that we use the graph theoretic method rather than the QuTiP solver for the pulsed SBC optimization because the graph theoretic method is computationally faster, allows for the use of a computational optimizer.
Finally, in all following pulsed cooling simulations results, it is assumed for simplicity that the quenching of the excited state is fast and requires a negligible amount of time. 
Thus, the recorded cooling time is only the time in which the cooling laser is interacting with the ion.

\subsection{Conversion to sideband ratio method}
\begin{figure}[t]
    \centering
    \includegraphics[width=\linewidth]{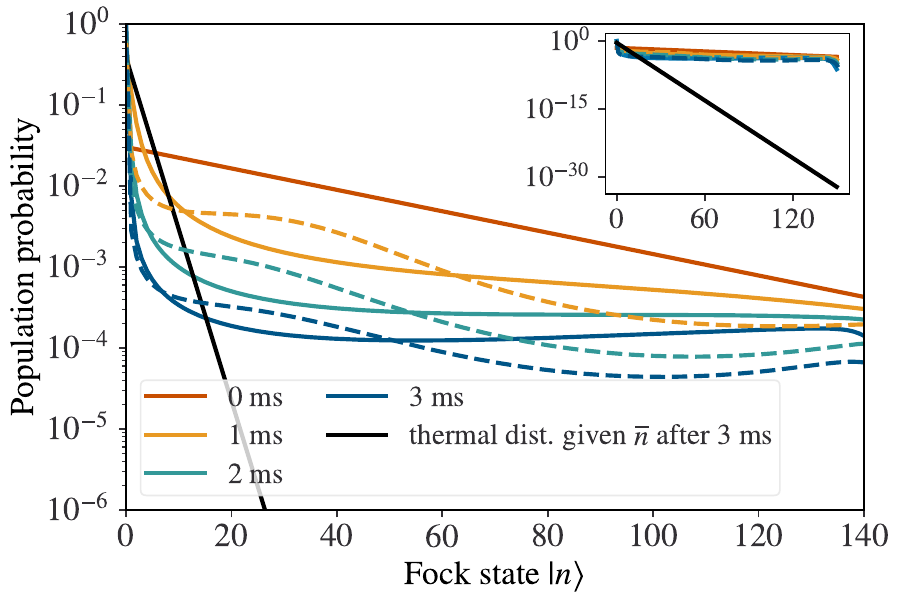}
    \caption{The Fock state population distributions for various cooling times for single order SBC. 
    The solid lines indicate continuous cooling while the dashed lines indicate pulsed cooling. 
    The black line shows the motional state after $3$\,ms of SBC if it were a thermal distribution as defined by the output value of $\overline{n}_{NS}$. 
    This shows how the assumption of the thermal distribution after SBC results in an overestimation of the low-lying motional states and an underestimation of the high order motional states. }
    \label{fig:postSBCFockDist}
\end{figure}
In our experiment, we measure the average phonon number of the ion with the commonly used ratio method in which one takes the ratio of the amplitudes of the first order red and blue sidebands:
\begin{equation}
    \overline{n}_{SBR}= \frac{P_r}{P_b - P_r} \; .
    \label{eq:SBRnbar}
\end{equation}
Here, $\overline{n}_{SBR}$ is the average phonon number as given by the sideband ratio method, and $P_r$ and $P_b$ are the amplitudes of the red and blue sidebands, respectfully.
Because the method relies on the difference between the red and blue sideband amplitudes, the error in the measurement becomes larger as the red and blue sidebands converge to comparable amplitudes, i.e., $\overline{n}_{SBR}	\gtrsim 1$.
However, it has been observed in previous studies \cite{Rasmusson2021, Vybornyi2023} that, even when $\overline{n}_{SBR} < 1$, the SBR method does not provide an accurate measurement of the true average phonon number: $\overline{n} = \sum_{n=0}^{n_{max}} P_n \langle n|a^{\dag}a|n\rangle $.
This is largely because the SBR method assumes a thermal distribution of motional states even after SBC \cite{PhysRevLett.62.403, Rasmusson2021, Roos1999}. 
Because of the assumption, from the post-SBC distribution of Fock state populations shown in Fig.~\ref{fig:postSBCFockDist}, we infer that the SBR method overestimates the population of the low-lying motional states while underestimating the population of highly excited states. 
However, taking the example of the 3\,ms post-SBC Fock state population distribution curves of Fig.~\ref{fig:postSBCFockDist}, the total amount of the population spread over Fock states $n>5$ is on the order of only $1\%$.
Therefore, while the SBR method does not provide an accurate measurement of $\overline{n}$, it may still serve as a reasonable check of near ground state preparation. 
\begin{figure}[t]
    \centering
    \includegraphics[width=\linewidth]{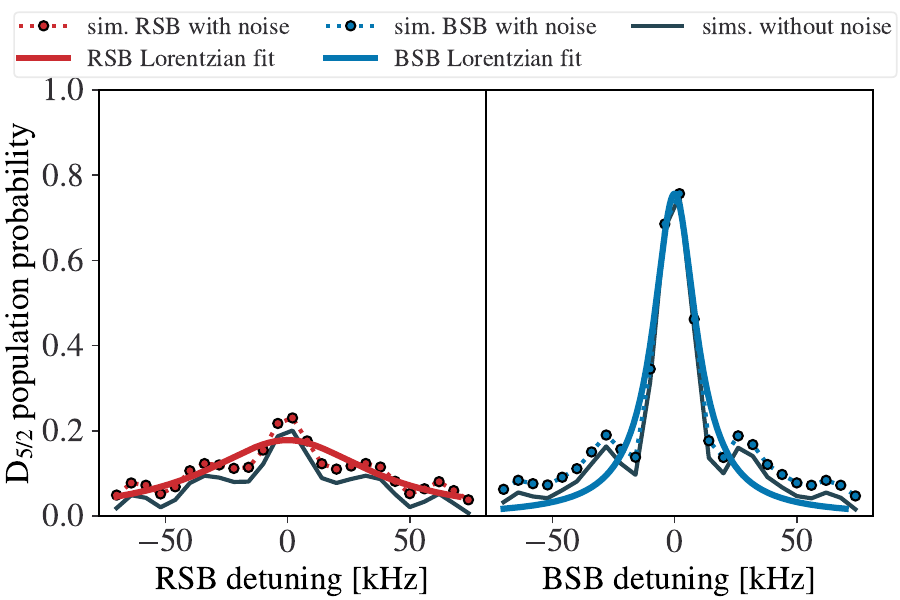}
    \caption{
    The amplitudes of the blue and red sidebands as a function of detuning from resonance with the respective sideband.  
    The solid gray lines give the sideband profiles without the \textit{post hoc} addition of noise while the red and blue points connected by dotted lines to guide the eye show the simulated sidebands with noise. 
    Red and blue solid lines show the Lorentzian fit to the simulated sidebands. 
    A Lorentzian fit was chosen as to not make any assumptions about the distribution of the motional state and was used to fit of the experimental data as well. 
    The fit of the heights of the blue versus red sideband allows one to extract a value for the average phonon number as demonstrated in Eq.~\ref{eq:SBRnbar}. }
    \label{fig:simulatedSidebands}
\end{figure}

\begin{figure}[t]
    \centering
    \includegraphics[width=\linewidth]{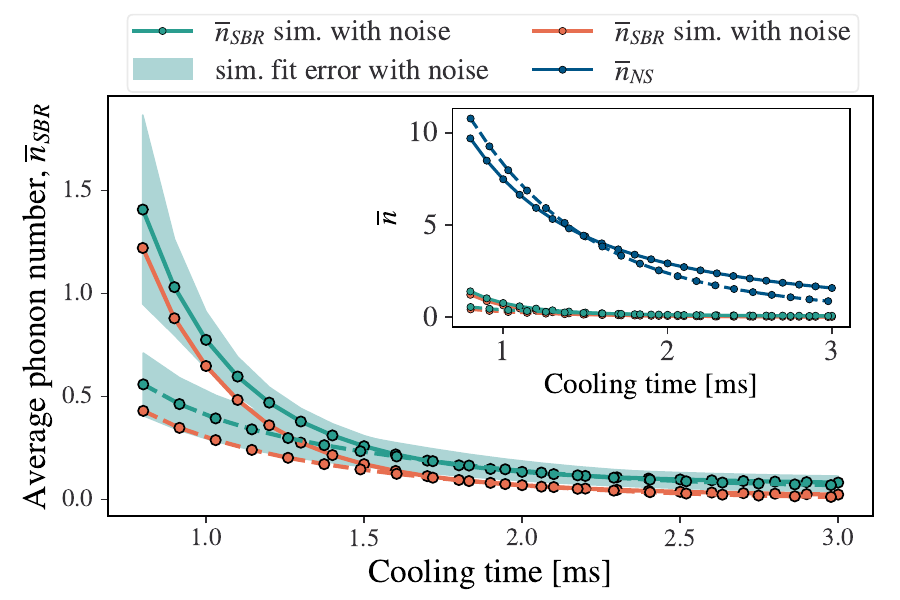}
    \caption{The cooling curves of optimized continuous (solid lines) and optimized pulsed (dashed) cooling as given by the numerical simulations and converted to $\overline{n}_{SBR}$. 
    The orange lines indicate the results without the \textit{post hoc} addition of noise while the green lines are with noise. 
    The inset plot also shows the results of the simulation, $\overline{n}_{NS}$, before the conversion to the SBR method. 
    This plot displays the large disparity between the true average of the phonon population distribution and the value of the average as determined by the SBR method.}
    \label{fig:simcoolingcurves}
\end{figure}

To make an accurate comparison between the simulations and experiments, we converted our numerical simulation results, $\overline{n}_{NS}$, to the equivalent result that one would measure under the SBR method of thermometry, $\overline{n}_{SBR}$. 
To do this, we use the motional state distribution output by either the master equation solver or the graph theoretic pulse simulator and simulate the resulting red and blue sideband spectral profiles as a function of laser detuning:
\begin{equation}
    P_{b/r}(\delta_{729}, T)=\varepsilon+\frac{1}{2} \sum_n^{n_{\max }} P_n \frac{\Omega_{n, n \pm 1}^2}{\Omega_{\textrm{eff}}^2}\left[1-\xi \cos \left(\Omega_{\textrm{eff}}T \right)\right] \; ,
    \label{eq:sincFunction}
\end{equation}
where 
\begin{equation}
    \Omega_{\textrm{eff}} = \sqrt{\Omega_{n, n \pm 1}^2+\delta_{729}^2} \; .
    \label{eq:effectiveRabi}
\end{equation}
Here, $P_n$ is the Fock state population for the state $|n\rangle$ as given by the output of the numerical simulation.
$\Omega_{n, n \pm 1}$ is the Rabi frequency of the blue ($+$) or red ($-$) sideband (Eq.~\ref{eq:RabiFreq}), $\delta_{729}$ is the detuning of the 729\,nm laser from resonance with a sideband transition, and $T$ is the interrogation time, which in this case corresponds to the ground state $\pi$-time of the blue sideband.
Because we want to make a fair comparison between the simulations and experiments, we must consider the impact of noise on the measurement. 
If one knows the major sources of noise in the experimental system, then one could include these sources in the simulated system evolution via the Lindbladian operators in the master equation. 
However, the graph theoretic approach does not currently have a means of modeling the effect of noise. 
Therefore, we include noise in our simulations \textit{post hoc} via an offset in Eq.~\ref{eq:sincFunction} to account for the experimental noise floor, $\varepsilon = 0.03$, and a decoherence parameter, $\xi =0.93$, which were determined by an experimental fit of the blue sideband Rabi flopping.
An example of these  simulated sidebands are shown in Fig.~\ref{fig:simulatedSidebands}. 
These sidebands are fit to a Lorentzian distribution as to not make any assumptions about the motional state distribution. 
Finally, the results of the conversion of the cooling simulation results to the SBR method are shown in Fig.~\ref{fig:simcoolingcurves}.

\section{\label{sec:exp} Experiment} 
In this section we give a brief overview of our experimental set-up.  
Additional information can be found in Ref. \cite{Qi2023}.
Our trap is a segmented blade Paul trap with a radius of $r_0 = 1.0$\,mm from the trap center to the electrodes. 
The rf drive frequency of $\omega_{rf}/2\pi = 19.2$\,MHz. 
A single trapped $^{40}$Ca$^+$ ion has typical secular frequencies of $(\omega_x, \omega_y, \omega_z)/2\pi = (1.12, 1.08, 0.415)$\,MHz.
At this axial (z-axis) secular frequency, the LD parameter for the 729-transition is $\eta\approx 0.15$.
A magnetic field with a magnitude of $6.2$\,G along the vertical axis is generated with permanent magnets to provide a quantization axis and split the otherwise degenerate Zeeman levels.

Neutral Ca produced by resistive heating is photoionized via a two-photon transition with 423\,nm and 379\,nm photons.
The trapped ion is then Doppler-cooled along the $S_{1/2} \rightarrow P_{1/2}$ transition with 397\,nm light and a repumping laser at 866\,nm to quench branching to the $D_{3/2}$ state out of the cooling cycle. 
The Doppler-cooling light is aligned to $45^{\circ}$ from the z-axis of the trap to allow projection onto all three axes of secular oscillation. 
The Doppler cooling limit is $0.53$\,mK, which for the axial secular frequency of $415$\,kHz corresponds to $26.8$ phonons, although the typical post-Doppler average phonon number is measured to be $30.5 \pm 2.2$ phonons as determined by a fit of the Rabi  flopping on the carrier transition. 
The ground Zeeman state $|S_{1/2},m_J=-1/2\rangle$ is initialized by optically pumping the $|S_{1/2},m_J=+1/2\rangle \leftrightarrow |D_{5/2},m_J=-3/2\rangle$ transition with $729$\,nm light that is aligned to the z-axis of the trap.
Then, the metastable excited state is quenched by exciting to $P_{3/2}$ with $854$\,nm light, which quickly decays back to the ground state. 
Finally, as previously discussed, sideband cooling is implemented on the $|S_{1/2},m_J=-1/2\rangle \leftrightarrow |D_{5/2},m_J=-5/2\rangle$ transition.
The $D_{5/2}$ is similarly quenched by 854\,nm light, which by selection rules populates only $|P_{3/2},m_J=-3/2\rangle$ and then decays only to $|S_{1/2},m_J=-1/2\rangle$ allowing for a closed SBC cycle.
The only means of leakage out of the cycle is due to off-resonant scattering to Zeeman levels with branching decay pathways.

To achieve a spectral linewidth narrow enough to excite the E2 transition from $S_{1/2} \leftrightarrow D_{5/2}$, the 729\,nm laser is first frequency-locked via the Pound-Drever-Hall (PDH) method to a high-finesse, low-thermal-expansion cavity with a linewidth of $15$\,kHz and a free spectral range of $1.5$\,GHz. 
While the PDH lock allows for the laser linewidth to be narrowed to approximately $10$\,kHz, it also produces ``servo-bumps" due to a finite locking bandwidth, which have previously proved detrimental to the experiment by exacerbating off-resonant scattering to the carrier transition.
Therefore, using the cavity also as a spectral filter, we use the transmitted light to seed a free running laser diode (FRLD).
The cavity filtering removes the servo-bumps, and the injection-lock to an FRLD amplifies the power from 3\,uW to 25\,mW while maintaining the spectral profile. 
Finally, to allow for on-demand optical power, the injection-locked FRLD is then used to seed a tapered amplifier, which amplifies the optical power to the range of $50-400$\,mW as desired for various other experiments. 

The frequency of the 854\,nm laser is also stabilized via a PDH lock to prevent frequency drifts. 
This laser is locked to a variable length transfer cavity with a linewidth of approximately $1$\,MHz and a free spectral range of $3$\,GHz.
An electro-optical modulator is used to shift the set-point of the PDH lock on the 854 laser so that the 854 frequency can be tuned to various values in experiments.
The cavity length is stabilized via PDH locking to an additional laser with a wavelength of 780\,nm, which is frequency stabilized via Doppler-free spectroscopy locking to a rubidium gas cell.

All SBC experiments presented here were implemented on the axial motional mode and were performed with the 729\,nm cooling laser tuned to a Rabi frequency of $\Omega^{01}_0/2\pi \equiv \Omega_{729}/2\pi = 50$\,kHz. 
The subsequent measurements of the sideband amplitude were performed with a Rabi frequency of $\Omega_{729}/2\pi = 70$\,kHz to reduce the impact of decoherence on the measurement. 
These correspond to $1.6$\,mW and $3.2$\,mW of optical power, respectively. 
For single order continuous SBC, the 854 repumping laser was set to $1.85$\,uW, which we estimate to be $\Omega^{12}_0/2\pi \equiv \Omega_{854}/2\pi =0.9$\,MHz.
This value was used as opposed to the optimal value of $1.34$\,MHz, because $0.9$\,MHz is the value at which the experimental calibration best matched the prediction of the numerical simulation as shown in Fig.~\ref{fig:exp854PowerCal} in the Appendix.
The same 854 power was used for multi-order continuous cooling, because $\Omega_{854}/2\pi=0.9$\,MHz was determined to be the optimum repumping Rabi frequency for both the second and first order red sidebands for multi-order cooling. 
The calculation of the optimum parameters for multi-order continuous SBC as well as the calibration of the 854 repumping laser Rabi frequency and detuning are discussed extensively in the Appendix. 
For pulsed SBC, because there are no strict power requirements for the 854 repumping laser, the optical power was raised to $50$\,uW to increase the repumping efficiency. 
At this 854 power, only $\sim 1$\,$\mu$s of repumping time is required to completely quench the $D_{5/2}$ state according to simulation.

\section{\label{sec:results} Results}
\begin{figure}[t]
    \centering
    \includegraphics[width=\linewidth]{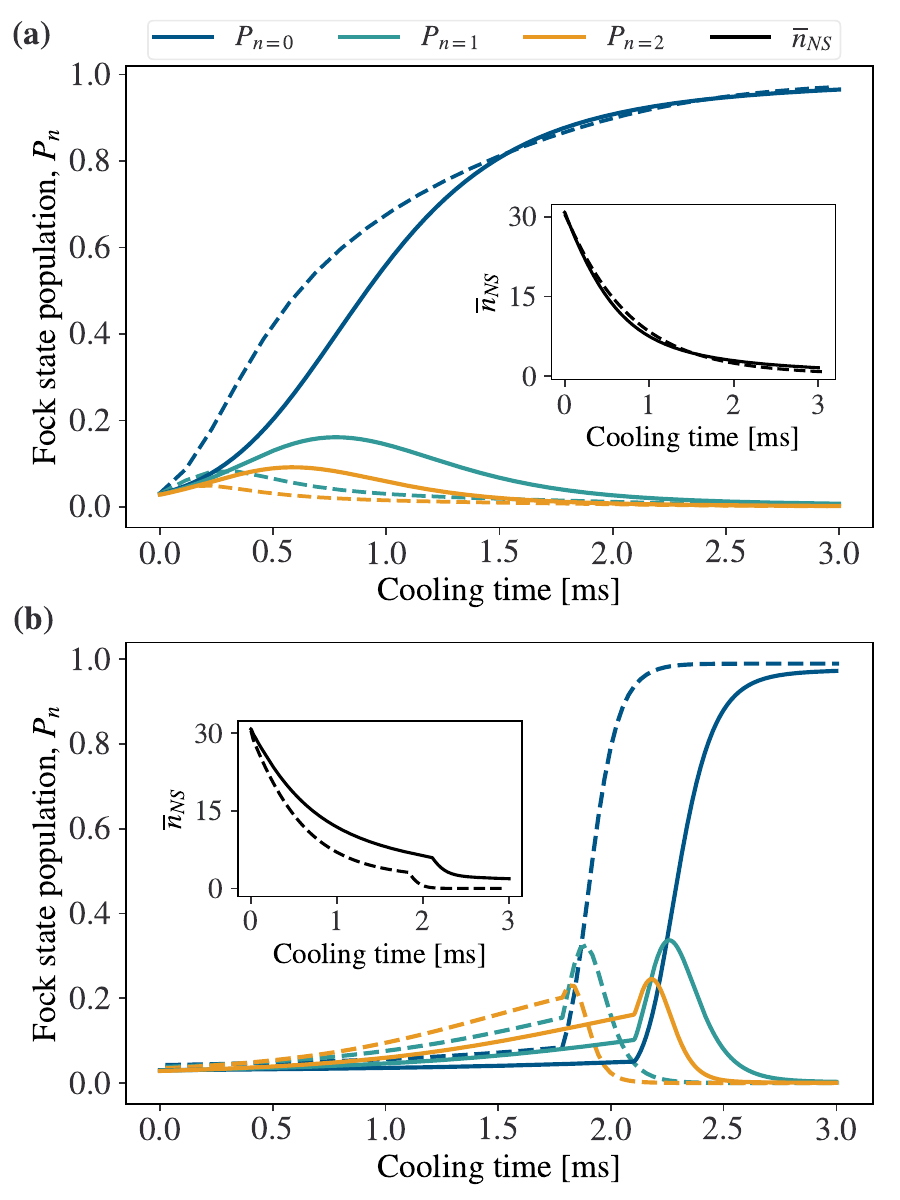}
    \caption{The evolution of the low-lying Fock states as a function of cooling time for (a) single order and (b) multi-order cooling all optimized for a total cooling time of $3$\,ms.
    The solid lines indicate continuous cooling while the dashed lines indicate pulsed cooling. 
    The insets of each plot show the evolution of the average phonon number, $\overline{n}_{NS}$.
    In plot (b), the switch from second order cooling to first order is seen when the slopes undergo a piece-wise shift after approximately $2$\,ms of cooling. }
    \label{fig:lowFockStates}
\end{figure}

The cooling curves generated through numerical simulation and converted to $\overline{n}_{SBR}$ both with (green) and without (orange) the estimated systematic noise for single order SBC can be seen in Fig.~\ref{fig:simcoolingcurves}.
Here, the solid lines indicate continuous cooling while the dashed lines indicate pulsed cooling.
In this continuous cooling simulation, the repumping laser detuning was set to $\delta_{854}=0$\,MHz and the Rabi frequency was optimized to a value of $\Omega_{854}/2\pi = 1.34$\,MHz. 
Meanwhile, in the pulsed cooling simulation, the duration of cooling laser pulses was optimized to a value of $\tau = 22.90$\,$\mu$s.   
The inset of Fig.~\ref{fig:simcoolingcurves} shows two things: (1) a comparison of the average phonon number $\overline{n}_{NS}$ (dark blue lines) for continuous versus pulsed SBC and (2) a demonstration of the drastic difference between $\overline{n}_{NS}$ and the average phonon number that one would measure in experiment under the sideband ratio method $\overline{n}_{SBR}$.
While the $\overline{n}_{NS}$ cooling curves in the inset show that single order continuous and pulsed cooling have comparable cooling rates, the $\overline{n}_{SBR}$ cooling curves of the main plot indicate otherwise. 
This conflicting narrative can be better understood by the results shown in Fig.~\ref{fig:lowFockStates}a. 
Here, this figure shows that pulsed cooling (dashed lines) more efficiently populates the low-lying Fock states at short cooling times compared to continuous cooling, but as cooling progresses the two methods arrive at the ground state concurrently. 
This explains the difference in the cooling rates of the $\overline{n}_{SBR}$ cooling curves in Fig.~\ref{fig:simcoolingcurves}: the set value of the repumping laser Rabi frequency was chosen to most efficient populate the ground state in the long time scale. 
A different value could have been chosen to efficiently populate the ground state in the short time scale, but it would have stalled at longer cooling times (see Fig.~\ref{fig:minNbar_maxPn=0}). 
Since the SBR method of measuring $\overline{n}$ over-weights the population of the low-lying motional states, this initial slow population is reflected in the $\overline{n}_{SBR}$ cooling curve as a lower cooling rate. 
However, in the limit of long cooling times as the system approaches the motional ground state, we conclude that the two single order techniques perform equally well. 

Fig.~\ref{fig:lowFockStates}b shows the multi-order simulation results in which cooling is performed on the second order RSB to start and is followed by cooling on the first order.
The switch between the two sideband orders can be observed by the piece-wise change in the slope. 
In the case of continuous SBC, the Rabi frequency of both the second and first order cooling stages were optimized as well as the ratio of time spent on the second order versus the first order RSB. 
The second order RSB was optimized for minimizing $\overline{n}$ rather than maximizing $P_{n=0}$ in order to more effectively remove population from highly excited motional states and because simulations showed the second order RSB to be ineffective at populating the motional ground state (see Fig.~\ref{fig:minNbar_maxPn=0_RSB2}). 
The optimization of multi-order continuous cooling is further described in the Appendix. 
For multi-order pulsed SBC, the pulse times were optimized for both second and first order RSBs, and the relative number of second order versus first order pulses were optimized as well. 
Note that the ratio of cooling time spent on the first order to second order is different for continuous cooling compared to pulsed cooling. 
From the results of Fig.~\ref{fig:lowFockStates}b, we observe that optimized multi-order pulsed SBC both more efficiently reduces the average phonon number and more quickly populates the motional ground state than optimized multi-order continuous SBC.

\begin{figure}[t]
    \centering
    \includegraphics[width=\linewidth]{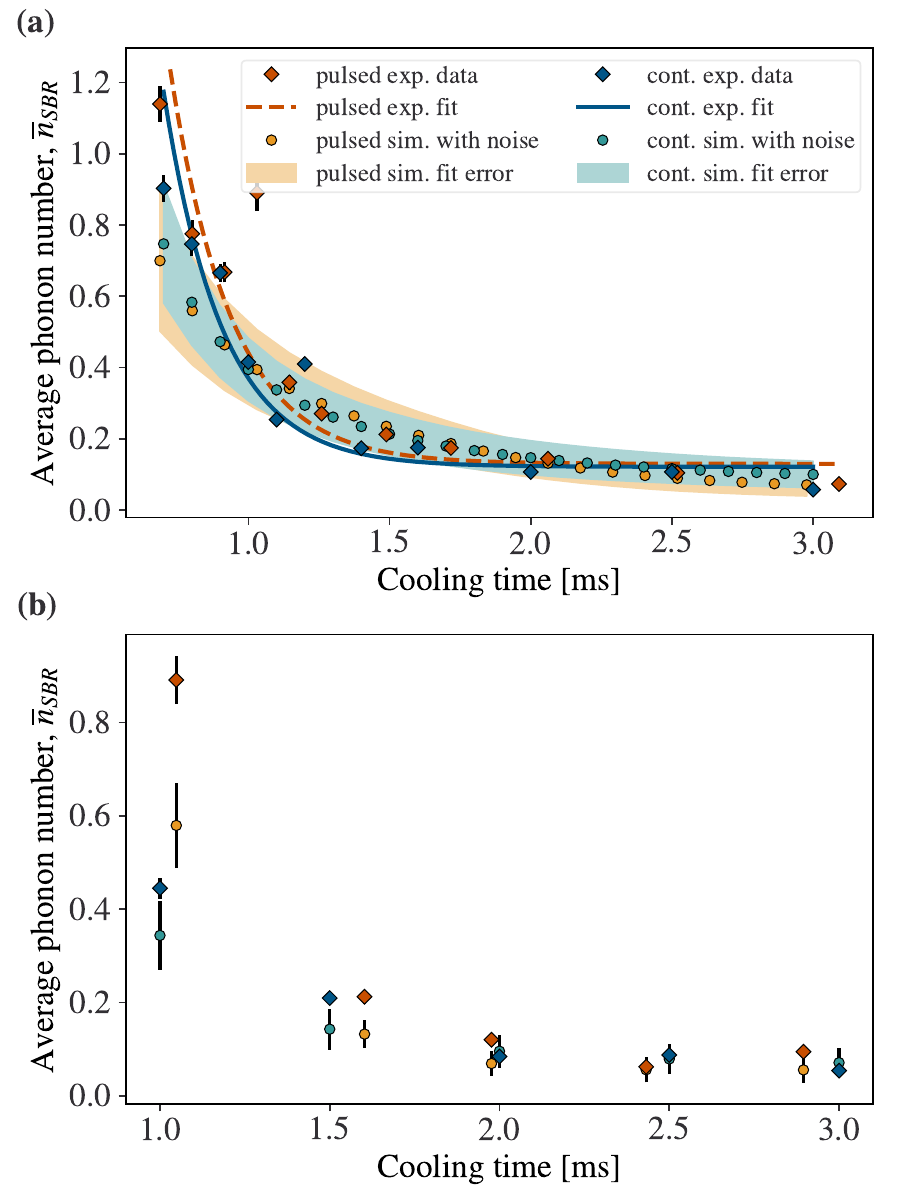}
    \caption{The experimental results (diamond markers) plotted along side the simulation results with systematic noise (circles) for (a) single order and (b) multi-order SBC. 
    The solid exponential decay fit lines indicate continuous cooling while the dashed fit lines indicate pulsed cooling. 
    The single order sideband results in (a) were optimized for the total cooling time of 3\,ms.
    The points show the temporal evolution of $\overline{n}_{SBR}$. 
    The multi-order sideband results (b) were optimized for its corresponding cooling time. }
    \label{fig:expResults}
\end{figure}

To confirm the validity of the simulation models, we measured $\overline{n}_{SBR}$ as a function of cooling time with a single trapped Ca$^+$ ion.  
Plots (a) and (b) of Fig.~\ref{fig:expResults} show the results of the single order and multi-order experiments, respectively, and how they compare to the numerical models under the same parameters.
In the single order SBC experimental results of Fig.~\ref{fig:expResults}a, we find good agreement with the numerical simulations.
Here, the experimental parameters were chosen based on the total cooling time of $3$\,ms as opposed to optimizing at every time step.
This approach was chosen because it reflects how cooling would be performed and thus how the motional state would evolve in a typical trapped ion experiment utilizing SBC. 
In this plot, the experimental datasets are fit to an exponential decay, $\overline{n}(t)=\overline{n}_i e^{-\gamma t} + \overline{n}_f$, to guide the eye with the solid lines indicating continuous cooling and the dashed lines indicating pulsed cooling. 
Meanwhile, the simulation datasets are plotted with the error in the fit of the simulated sidebands as discussed in Section~\ref{sec:numericalSim}.C.
Additionally, as previously mentioned, for single order continuous SBC, a repumping laser Rabi frequency of approximately $0.9$\,MHz was used rather than the optimal value of $1.34$\,MHz because the Rabi frequency of $0.9$\,MHz best matched the experimental calibration.
The Rabi frequency calibration methods for continuous SBC are discussed extensively in the Appendix.

In the multi-order results of Fig.~\ref{fig:expResults}b, contrary to the approach used in the single order experiments, the cooling parameters were optimized independently for each time step. 
This was done for two reasons.  
First, the second order cooling stage does not populate the low-lying motional states effectively enough for $\overline{n}_{SBR}$ to be a reliable measure of temperature, which can be inferred from in Fig.~\ref{fig:lowFockStates}b. 
Second, the optimization via numerical simulation produced different ratios for the time spent on second order RSB versus the first order. 
Even with independent optimizations for each time step, we find that the experimental results in Fig.~\ref{fig:expResults}b align well with the computational results. 
The good agreement between the simulation and experiment, even with the limitations of the ratio method of thermometry, grants credibility to the aspects of the simulations that are not as easily testable in experiment such as the evolution of the Fock state population distribution with respect to cooling time. 

\section{\label{sec:discuss} Discussion}
Despite its small advantage in cooling efficiency over continuous cooling, there are intrinsic disadvantages to pulsed SBC.
Firstly, the need to generate pulses as opposed to simply switching on a continuous-wave laser adds experimental complexity that can introduce problems with experimental control.
Secondly, as previously mentioned, the ``cooling time" for pulsed SBC refers to the time during which the cooling laser is interacting with the ion (i.e., the running sum of the duration of pulses) and does not include the time required to quench the metastable excited state. 
Though, this time can be made small compared to the duration of pulses with increased optical power and proper calibration of the repumping laser resonance.
In our experiments, with the repumping laser power set to $50$\,$\mu$W, the time to completely quench the excited state is approximately $1$\,$\mu$s.
However, if the repumping laser power is set to the same value used in the continuous SBC experiments, then the time required to quench the metastable excited state become comparable to the cooling pulse time. 
The drawback to the increased optical power of the repumping laser, however, is that it can lead to increased off-resonance coupling to other Zeeman levels with branching pathways out of the cooling cycle, thus necessitating more frequent optical pumping of the electronic ground state.
Conversely, the small amount of repumping laser power required for efficient continuous cooling minimizes the off-resonant coupling to other Zeeman levels such that a closed cooling cycle can be maintained during optimized continuous SBC.
However, in spite of these drawbacks for pulsed cooling, the optimization and experimental calibration is easy if one makes use of the graph theoretic method \cite{Rasmusson2021}.
Comparing this to continuous cooling, far more effort is required to computationally optimize and experimentally calibrate the cooling parameters in order to implement efficient cooling. 
Though, the optimization and calibration of continuous cooling only needs to occur once, and this is now the standard method in our laboratory.

\section{\label{sec:conclude} Conclusion}
In this paper, we have performed a detailed comparison of the cooling efficiencies of continuous and pulsed SBC as well as cooling on single and multiple sideband orders.
We perform this comparison via both numerical simulations and experiments.
We assessed the performance of the various techniques by observing the rate of population of the motional ground state as well as the rate at which the average phonon number approaches zero.
We started by reviewing the fundamentals of SBC and detailing the differences between continuous and pulsed cooling.
We then described the computational methods by which each cooling technique was optimized, and we reported our finding that, for continuous SBC, one can choose different parameters to optimize for either maximization of $P_{n=0}$ or minimization of $\overline{n}$.
We also detailed the inaccuracies of the sideband ratio method as a means of measuring of $\overline{n}$ but justified its use as a check of near ground state preparation. 
We then outlined the conditions of the experimental set-up and presented the results of the numerical simulation and the subsequent experiments.
The results of our investigation show that pulsed and continuous cooling perform roughly the same for single order SBC on the first order RSB.
However, for multi-order SBC, we found that optimized pulsed cooling outperformed optimized continuous cooling in both the rate of population of the ground state and the reduction of total $\overline{n}$. 
Though, the difference in cooling efficiency is not great enough to warrant a claim of general superiority of one method over the other. 
Finally, we verified the validity of the computational results by demonstrating good agreement with experiment.

\section{\label{sec:acknowledge} Acknowledgements}
This work is supported by the Army Research Office (W911NF-21-1-0346) and the Department of Energy Quantum Systems Accelerator project. 
We thank Geert Vrijsen for providing the initial body of the QuTiP simulation code and Swarnadeep Majumder for helpful conversations.

\section{\label{sec:Appendix} Appendix}

\subsection{Lindbladian operators in continuous cooling simulation}
\begin{table}[H]
    \centering
    \begin{tabular}{|c|c|c|c|} \hline 
         $L_{0}$ & $L_1$ & $L_2$ & $L_3$ \\ \hline 
         $\sqrt{A_1}\sigma_-$ & $\sqrt{b_{2 \rightarrow 1} A_2}\sigma_-$ & $\sqrt{b_{2 \rightarrow 0} A_2}\sigma_-$ & $\sqrt{R}(a^{\dag}+a)$\\ \hline 
          $1.3\sigma_-$&  $3.2\times 10^3 \sigma_-$&  $12.2 \times  10^3 \sigma_-$& $0(a^{\dag}+a)$\\ \hline
    \end{tabular}
    \caption{Lindbladian operators for the three-level system in Ca$^+$ used in numerical simulation of continuous SBC.
    The units of the given values of $L_i$ are in $\sqrt{\textrm{Hz}}$.}
    \label{tab:LindbladOperatorTable}
\end{table}

Here we provide the Lindbladian operators used in our numerical simulations of continuous SBC. 
These are specific to the three-level system in Ca$^+$ but are not unique to our system. 
The $A_i$ values for the states $|i\rangle$ are the sum of the Einstein coefficients ($\sum_f A_{if}$) as given by the NIST database \cite{NIST_ASD}, and the values of $b_{2 \rightarrow f}$ are the branching ratios ($b_{i \rightarrow f} =A_{if}/A_i)$ of the auxiliary excited state $|2\rangle$ to $|0\rangle$ or $|1\rangle$. 
We ignore branching to the D$_{3/2}$ to limit the  simulation to three levels. 
This is justified by the fact that the branching to this state is small (0.007), and in experiment this state is constantly repumped by the 866\,nm laser thereby equating to a decay to the S$_{1/2}$. 
We normalize the branching ratios to account for this elimination.
Moreover, due to our low single ion heating rate in our system of 11.4 phonons/s \cite{Qi2023}, we do not consider the effect of heating in the continuous SBC simulations in order to maintain consistency with the graph theoretic pulsed SBC simulations, which do not account for ion heating \cite{Rasmusson2021}, and we confirmed that this has a negligible effect on the results of simulation.

\subsection{Simulations in the Lamb-Dicke regime}
\begin{figure}[h]
    \centering
    \includegraphics[width=\linewidth]{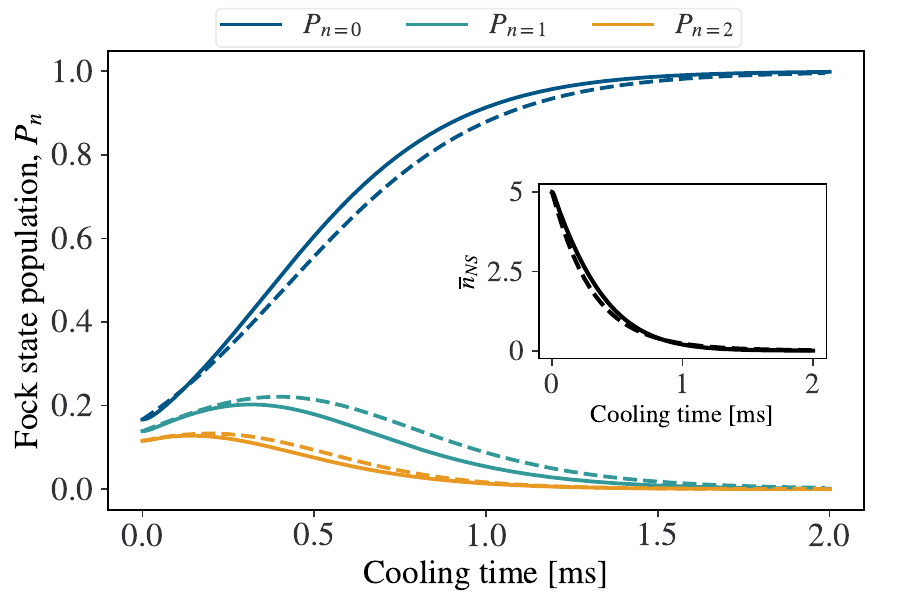}
    \caption{The low-lying Fock state populations for continuous (solid lines) and pulsed (dashed) SBC when the post-Doppler average phonon number is in the LD regime with $\overline{n}_{i}=5$ phonons.
    The inset shows the evolution of the average phonon number, $\overline{n}_{NS}$. 
    This demonstrates how even when the SBC process begins in the LD regime, the two SBC techniques perform similarly.}
    \label{fig:LowLyingFock_initnbar5}
\end{figure}

We now adjust our trap secular frequency in simulation to $\omega_{t}/2\pi=2.23$\,MHz such that the post-Doppler average phonon number is $\overline{n}=5$ phonons, which is well into the LD regime. 
The results of the simulations for these new conditions are shown in Fig.~\ref{fig:LowLyingFock_initnbar5}.
From this plot, we see that pulsed and continuous SBC still perform similarly even when ion motion begins in the LD regime. 
We do not test multi-order cooling because higher order sidebands are strongly suppressed in the LD regime. 

\begin{figure}[h]
    \centering
    \includegraphics[width=\linewidth]{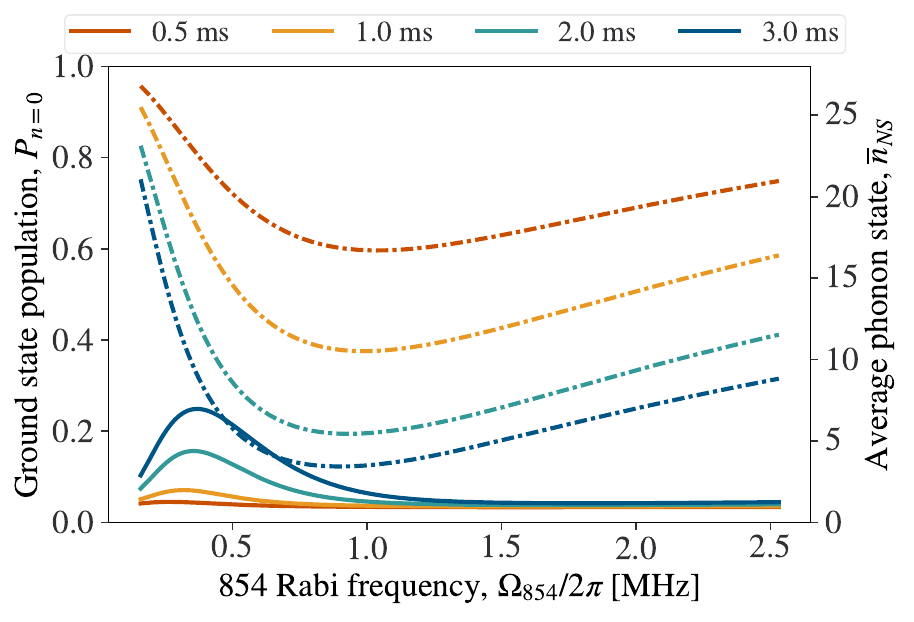}
    \caption{Ground motional state population (solid lines, left y-axis) and average phonon number evolution (dashed-dotted lines, right y-axis) as a function of cooling time on the second order RSB for continuous SBC at various cooling times with $\delta_{854}=0$\,MHz.
    This plot shows that the higher order sideband is ineffective at populating the ground motional state.}
    \label{fig:minNbar_maxPn=0_RSB2}
\end{figure}

\begin{figure}[t]
    \centering
    \includegraphics[width=\linewidth]{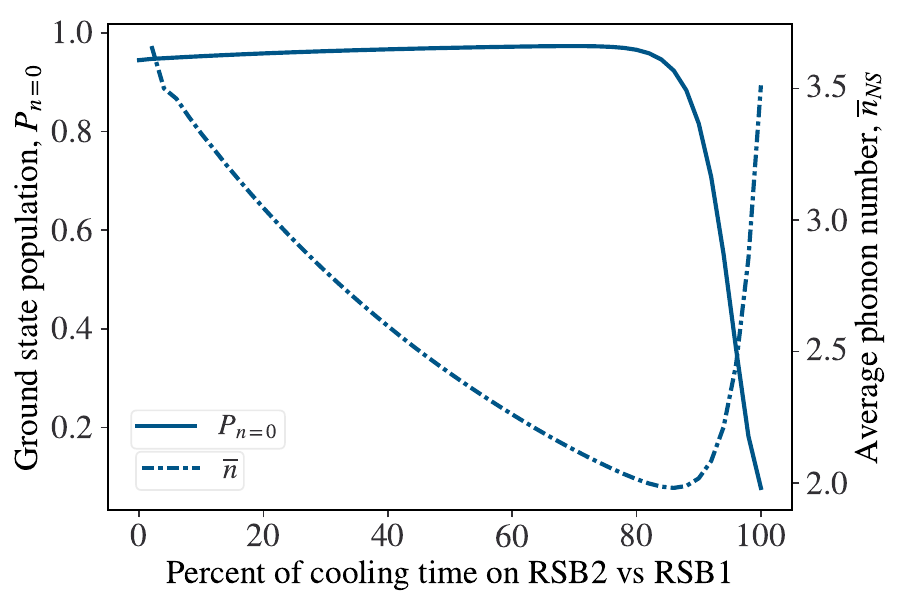}
    \caption{The effect of the varying the ratio of time spent cooling on the second order versus the first order RSB for continuous SBC for a total cooling time of 3\,ms. 
    Here, the solid line indicates the ground motional state population and corresponds to the left y-axis while the dash-dotted line indicates the average phonon number and corresponds to the right y-axis.}
    \label{fig:multiorderRatio}
\end{figure}

\subsection{Multi-order continuous SBC optimization}
Here, we discuss the additional computational effort required to optimize multi-order continuous cooling. 
The methods by which single order continuous SBC were optimized were discussed in Section \ref{sec:numericalSim}. 
Fig.~\ref{fig:minNbar_maxPn=0} shows how SBC on the first order RSB depends on the repumping laser Rabi frequency.
From this plot, since since the optimal Rabi frequency changes as a function of cooling time, we infer that the optimal value of $\Omega_{854}$ depends on the distribution of motional states. 
Moreover, this plot shows, through one's choice of $\Omega_{854}$, one can either optimize for maximizing the ground motional state population, $P_{n=0}$, or for minimizing the average phonon number, $\overline{n}$. 
When extending the optimization to higher order sidebands, one must consider the following: (1) for which quality each sideband will be optimized, high $P_{n=0}$ or low $\overline{n}$, (2) the amount of time spent on each sideband and the resulting motional state distribution as this will affect the parameters of the following sideband cooling stage, and (3) the optimal repumping Rabi frequency for each sideband given the initial conditions at the beginning of each cooling stage.  

We begin this process by analyzing the simulation results for cooling on the second order RSB given in Fig.~\ref{fig:minNbar_maxPn=0_RSB2}. 
We notice that the second order RSB does not effectively populate the ground motional state. 
One of the advantages of cooling at higher order RSBs is the remove of multiple phonons per cooling cycle. 
While this makes higher order sidebands effective at lowering the average phonon number, it does not effectively populate the ground state. 
Therefore, for the second order RSB, we choose to optimize for low average phonon number. 
Based on the optimization of multi-order pulsed cooling, we can estimate that the ratio of time spent cooling on the second order versus the first order will be approximately two to one.
Therefore, for a fixed total cooling time of $3$\,ms, we will spend approximately $2$\,ms on the second order RSB, which allows us to estimate that we will optimally cool to $\overline{n}\approx 6$ phonons if we set the repumping Rabi frequency to $0.9$\,MHz. 
Coincidentally, when starting the first order cooling stage at $\overline{n}= 6$ and optimizing for high $P_{n=0}$, the optimum repumping Rabi frequency for the first order RSB is also $0.9$\,MHz. 
Therefore, with these estimates of the optimal repumping Rabi frequencies, we can simulate cooling efficiency as a function of ratio of time spent on the second order RSB versus the first order. 
The results of this are shown in Fig.~\ref{fig:multiorderRatio}. 
From these simulation results, given our initial conditions and a total cooling time of $3$\,ms, we find that the optimum percent of time spent cooling on the second order versus the first order is 70\%.
The experimental implementation of multi-order continuous SBC is shown in Fig.~\ref{fig:expResults}b. 
For each time step in this plot, the ratio of time spent on the second order versus the first order was optimize. 
However, the repumping Rabi frequencies were held constant for each time step since the deviation of the optimal repumping Rabi frequency from $0.9$\,MHz is small with a difference of only $0.06$\,MHz and the experimental calibration of the 854 laser Rabi frequency does not allow for enough precision to warrant an adjustment of the laser power.
See below in Section \ref{sec:Appendix}.D and Fig.~\ref{fig:exp854PowerCal} for a description of the calibration of the 854 Rabi frequency.

\begin{figure}[t]
    \centering
    \includegraphics[width=\linewidth]{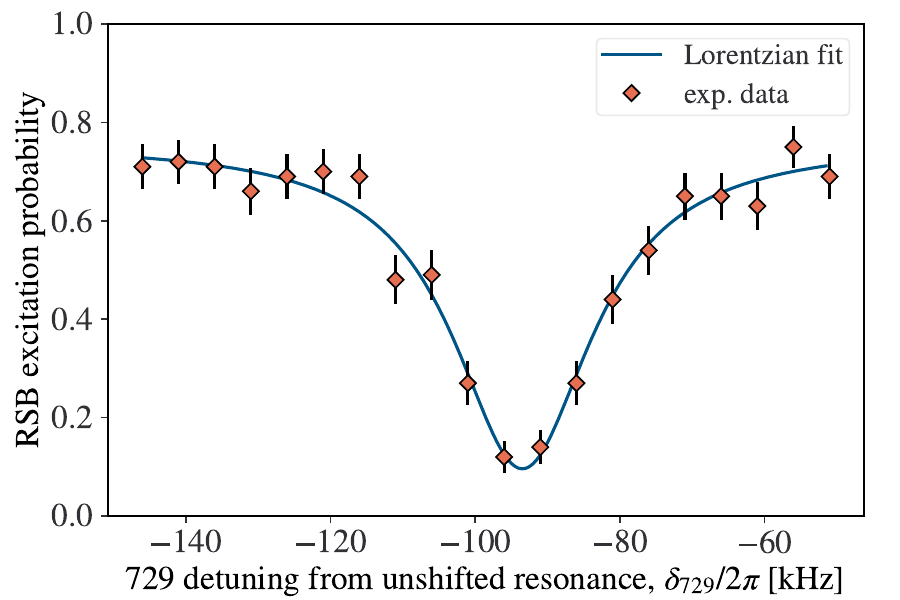}
    \caption{This figure shows the amplitude of the first order RSB after an attempt to continuously sideband cool as a means of finding the resonance frequency of the light shifted excited state. 
    During the scan, as the laser approaches resonance, the ion will be cooled and the height of the RSB will be reduced.}
    \label{fig:cscooling}
\end{figure}
\begin{figure}[t]
    \centering
    \includegraphics[width=\linewidth]{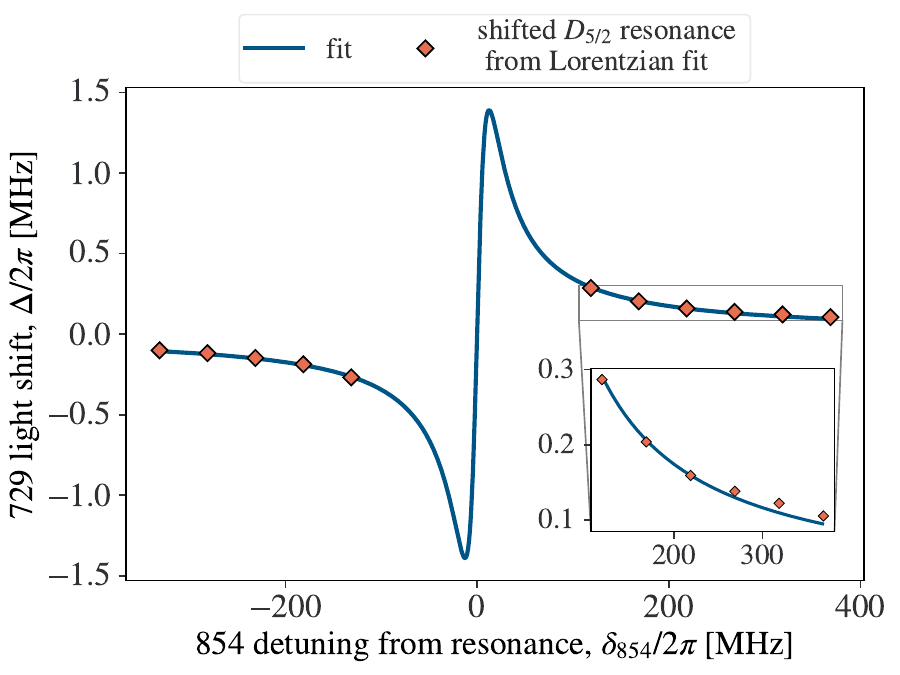}
    \caption{The light shift of the metastable excited state $D_{5/2}$ as a function of repumping laser detuning, which is described by Eq.~\ref{eq:LightShift}. 
    In this instance, the power of the repumping laser was set to 300\,uW. 
    Each point was extracted from a Lorentzian fit of the light shifted excited state as shown in Fig.~\ref{fig:cscooling}. 
    Additionally, each point is plotted with the error bar from the uncertainty in the fitted resonance value, but the error bars are smaller than the points. }
    \label{fig:StarkShift}
\end{figure}

\subsection{Light shift and 854 repumping laser calibration}
As previously discussed, efficient continuous SBC depends on precise calibration of the repumping laser (854) detuning and Rabi frequency. 
Figs. \ref{fig:2Dsim} and \ref{fig:minNbar_maxPn=0} show that the calibration of the 854 detuning and Rabi frequency must be within approximately $\pm 5$\,MHz and $\pm 0.5$\,MHz, respectively. 
Given the large natural linewidth of the E1 transition and the fact that scattering from the transition is not easily measured due to the short lifetime of the $P_{3/2}$ state, calibrating the 854 laser detuning and Rabi frequency to these narrow windows presents a challenging task. 
We approach this task first by measuring the magnitude of the light shift at various 854 optical powers and detunings. 
Then, we fine-tune the calibration by finding the parameters that allow for the best sideband cooling as measured by the SBR method of thermometry.

\begin{figure}[t]
    \centering
    \includegraphics[width=\linewidth]{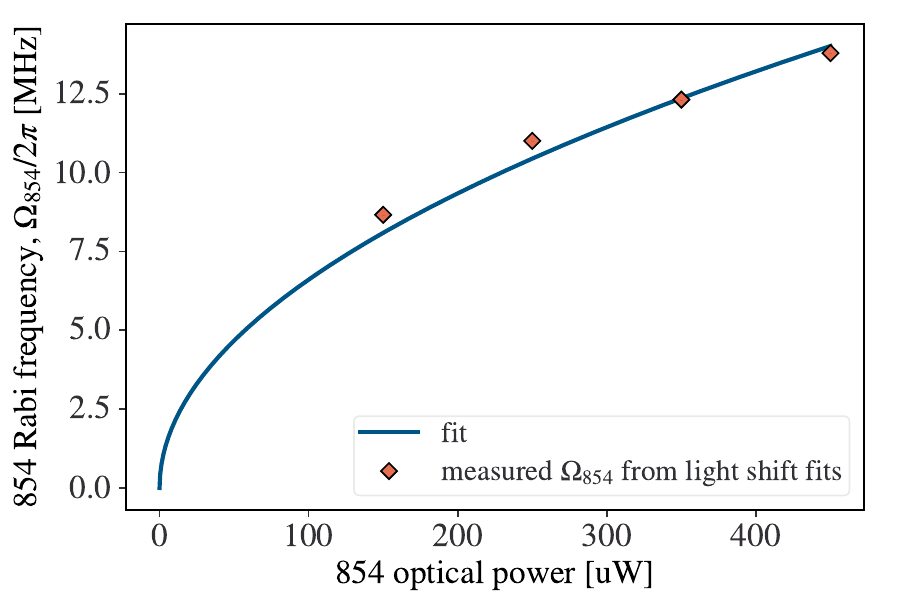}
    \caption{The conversion from measured 854 optical power to 854 Rabi frequency as determined by the fit of the light shifted excited state. 
    Each point corresponds to a fit of the light shift as a function of detuning for fixed optical power as shown in Fig.~\ref{fig:StarkShift}. 
    As in Fig.~\ref{fig:StarkShift}, each point is plotted with an error bar corresponding to the uncertainty in the Rabi frequency as determined by the fit, but the error bars are smaller than the points. }
    \label{fig:854RabiVsPower}
\end{figure}

\subsubsection{Light shift calibration}
In order to measure the magnitude of the light shift, $\Delta$, we attempt to sideband cool across a range of 729 frequencies and then measure the amplitude of the unshifted first order RSB. 
The 854 optical power is fixed and the frequency is set by shifting the set point of the PDH lock of the 854 laser with an electro-optical modulator as discussed in Section \ref{sec:exp}.
When the 729 frequency gets close to resonance with the light shifted RSB, the cooling is successful and the light-shifted resonance reveals itself as a reduction in the measured RSB amplitude. 
As we scan across the frequency range, this reduction in RSB height has the form of an inverted Lorentzian and can be seen in Fig.~\ref{fig:cscooling}.
From this Lorentzian fit, we can extract the point of resonance with the light shifted $D_{5/2}$.
This scan searching for the light shifted RSB resonance is repeated for multiple values of the 854 frequency with fixed 854 optical power.
We can then fit these data to the expression for the light shift magnitude given in Eq.~\ref{eq:LightShift}.
This fit is shown in Fig.~\ref{fig:StarkShift}.
From this fit of the light shift as a function of repumping laser detuning, we can extract values for the resonance frequency and the Rabi frequency that corresponds to the applied optical power. 
This process of measuring the light shifted $D_{5/2}$ resonance at various 854 frequencies and fitting the points to the light shift as a function of detuning is then repeated for multiple values of 854 optical power. 
This provides a set of measured values of the 854 resonance with a standard deviation of $13.04$\,MHz and a weighted average fit error of $\pm 0.22$\,kHz. 
Additionally, each light shift fit gives a measurement of the Rabi frequency associated with the applied 854 optical power. 
These Rabi frequencies are plotted in Fig.~\ref{fig:854RabiVsPower} and are fit using the square root proportional relationship between the Rabi frequency and optical power: $\Omega_0 = \alpha\sqrt{P}$, where $\alpha$ is a constant depending on various experimental parameters. 
Using this fit, we can convert our 854 Rabi frequencies found from simulation to experimental optical power.

\begin{figure}[t]
    \centering
    \includegraphics[width=\linewidth]{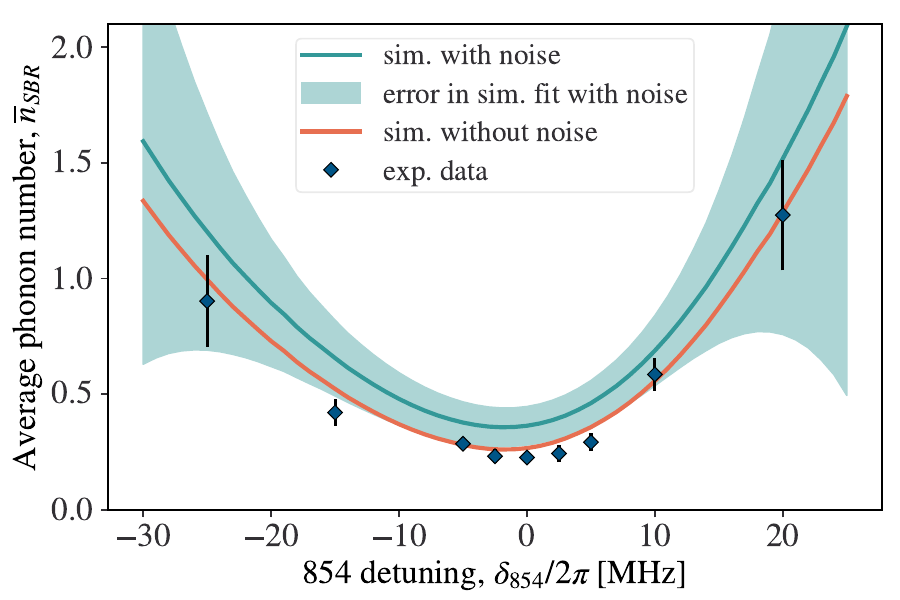}
    \caption{The calibration of the repumping laser detuning. 
    For a fixed cooling time of 1\,ms we vary the 854 frequency and find the value that cools closest to the ground state. 
    This is plotted with the simulation results converted to the SBR method both with and without the \textit{post hoc} noise model. 
    We find good agreement between the simulations and experiment.
    The zero of the x-axis is set with respect to the measured resonance of from the fits of the light shift.}
    \label{fig:exp854FreqCal}
\end{figure}

\begin{figure}[t]
    \centering
    \includegraphics[width=\linewidth]{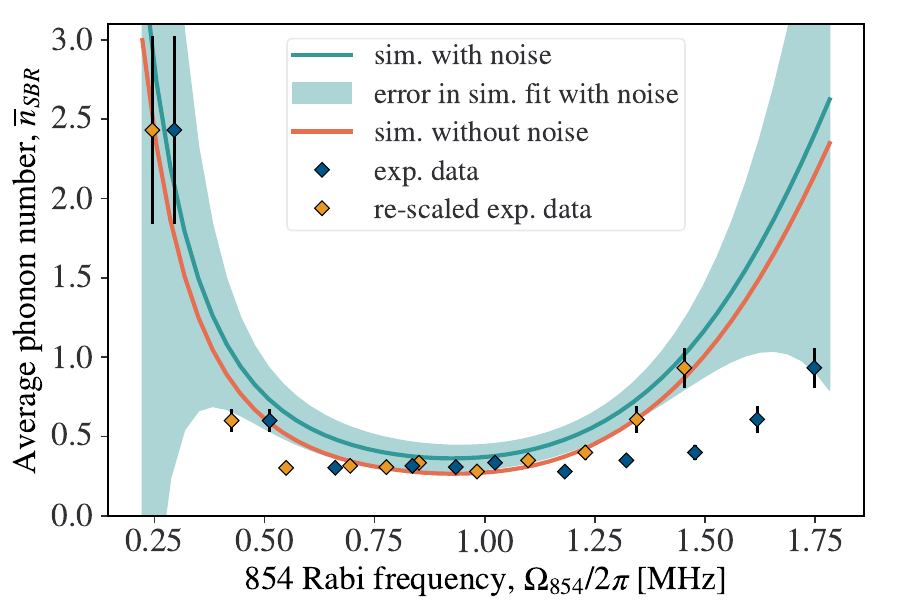}
    \caption{The calibration of the repumping laser Rabi frequency. 
    Just as was done for the calibration of the 854 detuning in Fig.~\ref{fig:exp854FreqCal}, for a fixed cooling time of 1\,ms the 854 power was varied to find the value that provides the best cooling. 
    The results of the simulations are also shown. 
    The measured optical power for each point was converted to Rabi frequency according to the fit in Fig.~\ref{fig:854RabiVsPower}. 
    Upon finding a disagreement between the simulation and this scaling of the experimental data, the power-to-Rabi-frequency conversion was re-scaled to better align the simulated results. 
    Both power-to-Rabi-frequency conversions are shown here. }
    \label{fig:exp854PowerCal}
\end{figure}

\subsubsection{Repumping laser calibration}
To further calibrate the 854 detuning and Rabi frequency, we sideband cool the ion for $1$\,ms to partially populate the ground state, and we vary the 854 detuning and power to find the values that minimize $\overline{n}_{SBR}$.  
We then compare these experimental results to numerical simulations. 
Fig.~\ref{fig:exp854FreqCal} shows the results of the 854 detuning calibration.
Here, the x-axis is defined with respect to the resonance as found by the light shift measurements detailed in the previous subsection.
We can see that the experiments and the simulated effect of repumping laser detuning on the SBC efficiency match very well. 
Fig.~\ref{fig:exp854PowerCal} shows the calibration the 854 Rabi frequency. 
The x-positions of the dark blue data points are determined by the power-to-Rabi-frequency conversion as determine by the light shift measurements found in the previous subsection. 
In this figure, one can see the mismatch between this conversion and the expected cooling efficiency from the numerical models. 
The yellow points are the same experimental data but with the $\Omega_{854}\propto \sqrt{P}$ relationship re-scaled to better align with the simulation. 
Since the latter approach injects bias into the Rabi frequency calibration, we set our 854 power for experiments to be the point where $\overline{n}_{SBR}$ is approximately minimized for both Rabi frequency scales: $1.85$\,uW.
We estimate this to be approximately $\Omega_{854}/2\pi= 0.9$\,MHz.

\bibliography{SidebandCooling}
\end{document}